# Thermal conductivity for p-(Bi, Sb)$_2$Te$_3$ films of topological insulators

L N Lukyanova[1], Yu A Boikov[1], O A Usov[1], V A Danilov[1], I V Makarenko[1], V N Petrov[1]

[1]Ioffe Institute, Russian Academy of Sciences, 26 Politekhnicheskaya, St Petersburg 194021, Russia

E-mail: lidia.lukyanova@mail.ioffe.ru

**Abstract**

The temperature dependences of the total, crystal lattice and electronic thermal conductivities were investigated in films of topological insulators p-Bi$_{0.5}$Sb$_{1.5}$Te$_3$ and p-Bi$_2$Te$_3$ formed by discrete and thermal evaporation methods. The largest decrease in the lattice thermal conductivity owing to the scattering of long-wavelength phonons on the grain interfaces was observed in the films of solid solutions p-Bi$_{0.5}$Sb$_{1.5}$Te$_3$ deposited by discrete evaporation on the amorphous substrates of polyimide without thermal treatment. It is shown that in the p-Bi$_{0.5}$Sb$_{1.5}$Te$_3$ films with low thermal conductivity the energy dependence of the relaxation time is enhanced, which is specific for the topological insulators. The electronic thermal conductivity was determined taking into account the effective scattering parameter in the relaxation time approximation versus energy in the Lorentz number calculations. The observed increase of the electronic thermal conductivity within the temperature range of 40 - 80 K is related to the weakening of the electrical conductivity temperature dependence and is determined by the increase in the effective scattering parameter at low temperatures due to the effect of scattering on the point antisite and impurity defects. A correlation was established between the thermal conductivity and features of the morphology of the interlayer surface (0001) in the studied films.

## 1. Introduction

Thermoelectrics based on bismuth and antimony chalcogenides are widely known as effective materials, which possess properties optimal for the temperature range of 100 - 500 K depending on the composition and charge carrier concentration [1, 2]. Currently, bismuth and antimony chalcogenides attracted much attention as promising topological insulators (TIs), in which topological surface states (TSS) arise due to inversion of the energy gap edges caused by strong spin-orbit interaction. In addition, the bulk becomes an insulator, and the electrons on the surface acquire unusual spin-momentum locked TSS with linear dispersion and spin polarization specific for the Dirac fermions [3, 4]. Such topological phenomena expand the application possibilities of TIs in various fields of physics [5 - 8] including the thermoelectricity [9, 10].

An enhancement of thermoelectric performance in $Bi_2Te_3$ based films of TIs is associated with an increase in the energy dependence of the spectral distribution of the mean free paths of both phonons and electrons [9, 10]. The estimations of the energy dependence of the mean free path of the electrons in TIs have shown that the electrons possess wider spectrum than phonons. The effect of electron energy filtering in TIs [11, 12] associated with an increase of the Seebeck coefficient due to energy dependent scattering of the charge carriers at the grain interfaces in the films [13, 14].

The strong energy dependence of the mean free path of phonons in a narrow energy range, which is significantly less than the electron energy range, is accounted for by the intensive phonons scattering on an interface between two grains in a polycrystalline material that results in decrease of the crystal lattice thermal conductivity in chalcogenide films [9, 14 - 16]. The effect of phonon scattering on grain interfaces on reduction of the lattice thermal conductivity in solid solutions based on bismuth telluride is associated with features of the phonon spectrum, in which the acoustic phonons with low-frequency and long wavelength have the mean free path values larger than high-frequency ones [17].

The largest heat transfer in the considered films is determined by long-wavelength phonons, which mainly affect the decrease in the lattice thermal conductivity $\kappa_L$ due to enhance their scattering. However, at low temperatures up to the Debye temperature the decrease in $\kappa_L$ is determined mainly by phonons scattering on acceptor antisite defects and impurity atoms in solid solutions [18, 19]. Furthermore, an additional decrease in $\kappa_L$ occurs in the layered films of $Bi_2Te_3$ and its solid solutions when phonons are scattered on van der Waals interfaces in the superlattices with a period of about 2 nm consisting of two inverted quintiples between the Te (1) layers [20]. With increasing of temperature, the effect of scattering on the $\kappa_L$ by point defects is reduced while the scattering of phonons on the grain interfaces becomes dominant. But near the room temperature, the contribution of the phonon-phonon scattering noticeably increases [15, 21 - 23].

Specific feature of the topological thermoelectrics is an existence of the residual bulk electrical conductivity related to the presence of bulk defects [24, 25]. The reduction in bulk conductivity occurs as a result of the optimization of the thermoelectric composition owing to mutual compensation of contributions of acceptor and donor intrinsic defects providing the increase of TSS contribution to the total conductivity. However, in $Bi_2Te_3$-based films, the residual bulk conductivity cannot be completely eliminated. The listed properties of TIs that determine a decrease in thermal conductivity and an increase in the thermoelectric power coefficient (Seebeck coefficient) provide an enhance in the thermoelectric performance of chalcogenide films compared to bulk thermoelectrics in spite of a slight reduction in the electrical conductivity.

In this work, the effect of scattering mechanisms on total $\kappa$, lattice $\kappa_L$ and electronic $\kappa_e$ thermal conductivities was investigated, which associated with the peculiarities of the energy dependence of the relaxation time in the p-$Bi_{0.5}Sb_{1.5}Te_3$ and p-$Bi_2Te_3$ films obtained by discrete and thermal evaporation methods. The choice of formation technique and the film composition optimization allow to optimize the scattering mechanisms of phonons and electrons, which promote a decrease in the lattice thermal conductivity $\kappa_L$ due to the effect of the interlayer surface morphology, grain interfaces and van der Waals superlattice.

**2. Deposition technique and film structure**

The polycrystalline films of solid solutions p-(Bi,Sb)$_2$Te$_3$ and p-Bi$_2$Te$_3$ were obtained by the methods of discrete and thermal evaporation in vacuum. During the films formation by discrete evaporation, the initial material prepared in the form of a powder with a grain size of about 10 µm was passed by small portions into heated quartz crucible, where it instantly evaporated. Unlike the discrete evaporation during the thermal evaporation in vacuum there are two physical processes: the evaporation of the heated initial material and its deposition on the heated substrate [26]. The variation of temperature of the crucible and the substrate material of mica (muscovite) and polyimide showed that the optimal temperatures to obtain the films with specified composition are 800-850 ºC for the evaporator and 250-300 ºC for the substrate. The technological parameters of the formed p-Bi$_{0.5}$Sb$_{1.5}$Te$_3$ and p-Bi$_2$Te$_3$ films for thermal conductivity studies are given in table 1.

**Table 1**. The p-Bi$_{0.5}$Sb$_{1.5}$Te$_3$ and p-Bi$_2$Te$_3$ films formation technique parameters

| № | Formation technique | Substrate | Heat treatment | Thermoelectric power coefficient, µV K$^{-1}$ |
|---|---|---|---|---|
| | | *p*-Bi$_{0.5}$Sb$_{1.5}$Te$_3$ | | |
| 1 | discrete evaporation | polyimide | unannealed | 242 |
| 2 | discrete evaporation | polyimide | annealed | 215 |
| 3 | thermal evaporation | polyimide | unannealed | 200 |
| 4 | discrete evaporation | muscovite | annealed | 223 |
| | | *p*-Bi$_2$Te$_3$ | | |
| 5 | discrete evaporation | polyimide | unannealed | 234 |
| 6 | thermal evaporation | muscovite | unannealed | 203 |

To describe the structure of Bi$_2$Te$_3$ and its solid solutions, a primitive rhombohedral or hexagonal unit cell is used. While the *a* and *c* parameters of the hexagonal unit cell of the Bi$_2$Te$_3$ are 4.385 Å and 30.487 Å, respectively. The Te atoms possess either six Bi neighboring atoms or by three Bi and Te atoms. It allows to distinguish in the sequence of simple layers more complex formations consisting of five layers, which are called quintuples. In the considered hexagonal unit cell there are three such quintuples. The chemical bonds in the layers are mainly covalent with some degree of ionicity. The atomic layers of Te and Bi in the quintuple alternate in the sequence of (- Te(1) – Bi – Te(2) – Bi - Te(1) -). It is known that in the p-Bi$_{0.5}$Sb$_{1.5}$Te$_3$ solid solutions, the Sb atoms substitute Bi. In these materials, an intrinsic antisite point defects on the sites of tellurium Bi$_{Te}$ and impurity defects caused by Sb → Bi atoms substitutions in solid solutions are revealed. The quintuples boundaries are the interlayer van der Waals surfaces or the cleavage planes (0001). The quintuples are bonded by the weak van der Waals forces, which provides a slight exfoliation of the crystal along the (0001) planes perpendicular to the crystallographic c-axis. The layered structure of the $Bi_2Te_3$ crystals and its solid solutions determines the significant anisotropy of the transport properties.

In the considered materials, an interlayer surface (0001) possesses the minimum value of free energy [27, 28]. It leads that the stable crystal seeds of the bismuth and antimony chalcogenides are formed with orientation predominantly along the c-axis perpendicular to the substrate plane even on substrates with large mismatch of the crystal lattice parameters. In the process of heat treatment at T = 390 ºC in the Ar atmosphere an intense selective evaporation of the Te atoms from the grain interfaces is occurred in addition to the secondary recrystallization. The depletion of the grain interfaces by tellurium affects the thermoelectric properties and in turn leads to an increase in the majority charge carrier concentration in the p-$Bi_{0.5}Sb_{1.5}Te_3$ and p-$Bi_2Te_3$ films [15]. It should be noted that according to the morphology studies of the interlayer surface (0001) of the p-$Bi_2Te_3$ films by conductive atomic force microscopy [16], an increase of local electrical conductivity in the region of the grain interfaces was observed.

## 3. Atomic force microscopy study of interlayer surface (0001) in the films

The investigation of the interlayer surface (0001) morphology in the p-$Bi_{0.5}Sb_{1.5}Te_3$ and p-$Bi_2Te_3$ films was carried out by the atomic force microscopy (AFM) in the semicontact mode.

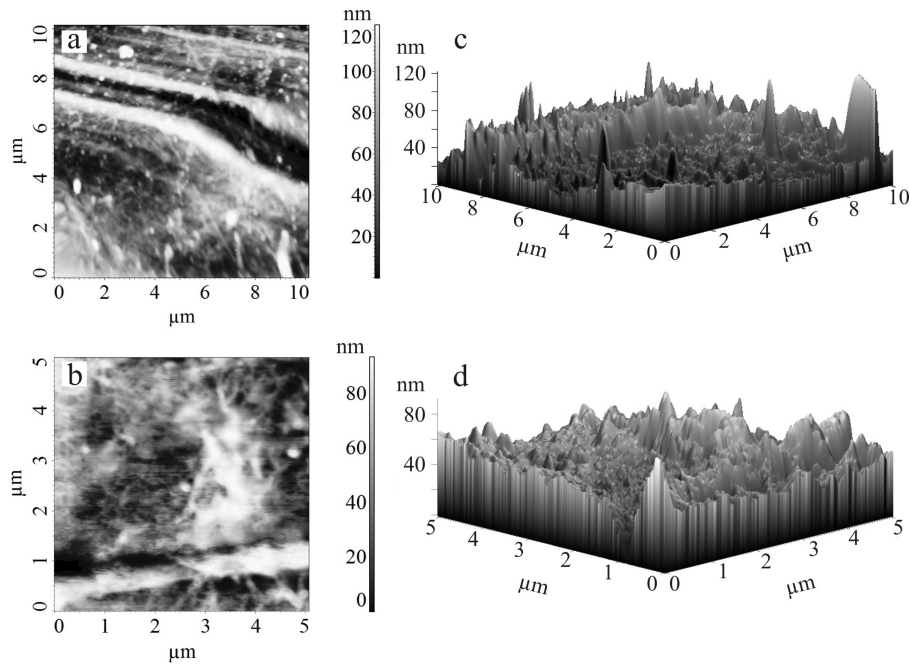

**Figure 1.** The (a, b) 2D and (c, d) 3D images of surface morphology (0001) of the unannealed p-$Bi_{0.5}Sb_{1.5}Te_3$ film obtained by discrete evaporation on polyimide substrate (a, c) and of the annealed one deposited on mica substrate (b, d).

Typical for all studied samples, two-dimensional (2D) and three-dimensional (3D) morphology images, the profiles and the histograms of nanofragment distribution on the surface (0001) depending on its height are shown in figure 1-3. As an example, the p-$Bi_{0.5}Sb_{1.5}Te_3$ films were analyzed (samples 1, 4, table 1). In these films relief of the surface (0001) was found to composed of separate nanofragments, islands, terraces consisting of coalescent islands, and rows containing dislocations (figure 1). The observed relief is formed by the diffusion processes and elastic stresses, which result in deformation of the interlayer surface during the films formation. Separate nanofragments arising on the (0001) surface are scattering centers for phonons leading to decrease in thermal conductivity of films based on chalcogenides of bismuth and antimony.

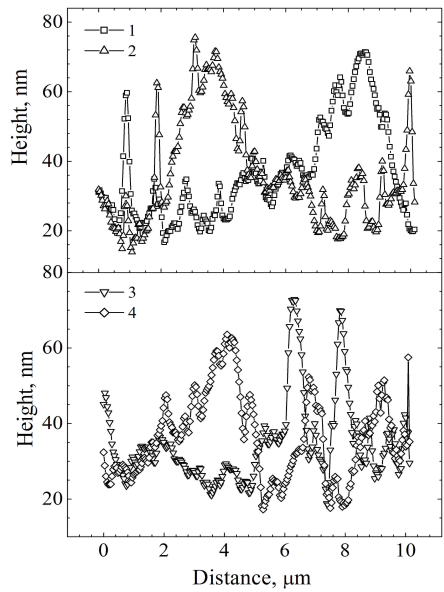

**Figure 2.** The profiles (1 - 4) of the surface morphology (0001) images of the p-$Bi_{0.5}Sb_{1.5}Te_3$ films (sample 1, table 1) obtained along arbitrary and mutually perpendicular directions in figure 1 a: 1, 2 and 3, 4. Average height of nanofragments on the surface (0001) in nm: 1 - 36.35; 2 - 36.24; 3 - 35.54; 4 - 36.49.

The average heights of nanofragments in the image of the surface morphology (0001) (figure 1 a) are about of 36 nm and have close values for profiles obtained on various surface regions (figure 2), which indicates the homogeneity of the relief of sample surface. More detailed information about the surface relief (0001) was found from the analysis of histograms (figure 3) determined from morphology images of the p-$Bi_{0.5}Sb_{1.5}Te_3$ films (samples 1, 4, table 1, figure 1).

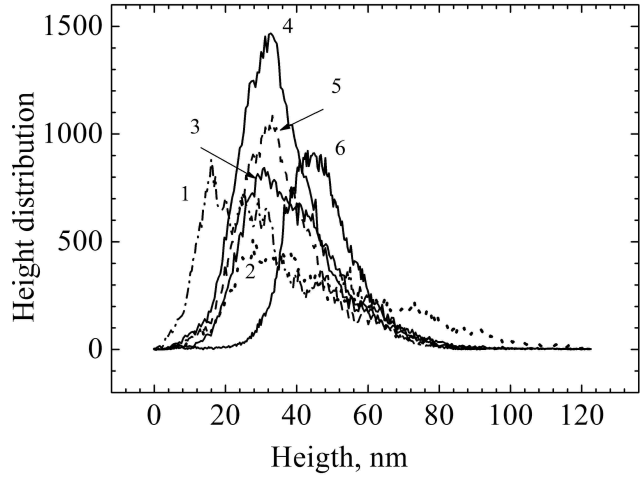

**Figure 3.** Height distribution of the nanofragments on the interlayer surface (0001) in the p-$Bi_{0.5}Sb_{1.5}Te_3$ films depending on the heights (1 - 3) in the annealed film obtained by discrete evaporation on mica substrate and (4 - 6) in the unannealed film deposited on polyimide substrate.

One can estimate the influence of annealing from the analysis of the nanofragments distribution on the interlayer surface (0001) depending on its height by the example of the p-$Bi_{0.5}Sb_{1.5}Te_3$ films obtained by discrete evaporation (figure 3). The number of nanofragments with sizes ranging from 16 to 28 nm (figure 3, curves 1 - 3) was maximal for p-$Bi_{0.5}Sb_{1.5}Te_3$ film subjected to annealing (table 1, sample 4). For the unannealed film, the dimensions of the most of nanofragments were from 31 to 45 nm (table 1, sample 1, figure 3, curves 4 - 6).

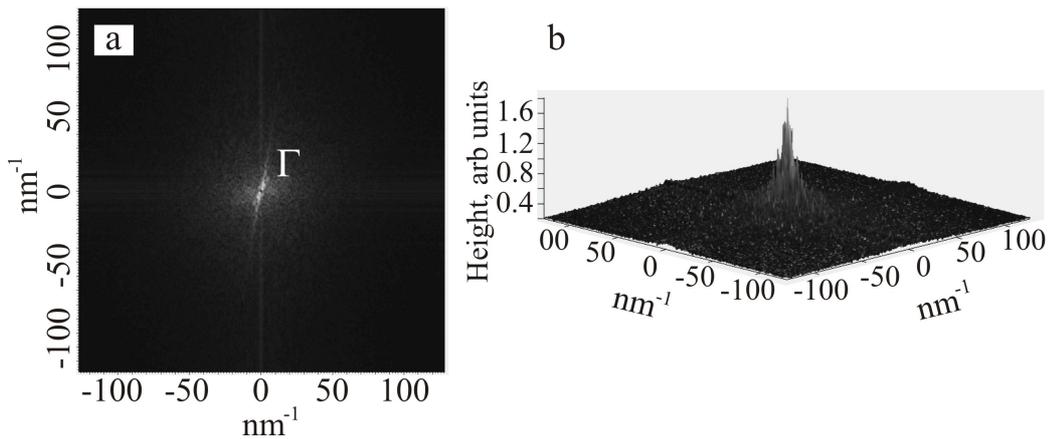

**Figure 4.** Fourier images of the surface morphology (0001) of the annealed p-$Bi_{0.5}Sb_{1.5}Te_3$ film obtained by discrete evaporation method on mica substrate. a - 2D and b - 3D Fourier images, respectively.

The average heights of nanofragments $R_a$ and the root mean square of height deviations of nanofragments $R_q$ (i.e. roughness) vary from 5.5 to 7 nm in the annealed p-$Bi_{0.5}Sb_{1.5}Te_3$ film, while in the unannealed film $R_a$ and $R_q$ increase to 10 - 15.85 nm. So, the influence of annealing results in both the decrease in the number of nanofragments of maximum sizes and the decrease of $R_a$ and $R_q$ values compared with the unannealed p-$Bi_{0.5}Sb_{1.5}Te_3$ film.

Fourier images of the film surface (0001) morphology for annealed p-$Bi_{0.5}Sb_{1.5}Te_3$ deposited by the discrete evaporation on a mica substrate were obtained using the Fast Fourier Transform (FFT). These images are intensities spectral distribution of two-dimensional reciprocal space being centered at the $\Gamma$ point of the Brillouin zone (figure 4 a, b). Friedel oscillations observed as intensity features of the spectral components in the vicinity of the $\Gamma$ point of the Brillouin zone are known to be specific for topological insulators and are associated with the interference of quasiparticle excitations of surface electrons on defects [29 - 31]. On these images, the spectral components are scaled in the vicinity of point $\Gamma$ (figure 4). The components of higher orders on Fourier images of the interlayer surface in solid solutions based on $Bi_2Te_3$ were found by scanning tunneling microscopy with atomic resolution [32].

The grain parameters were determined from the analysis of the surface morphology images of the p-$Bi_{0.5}Sb_{1.5}Te_3$ films obtained by discrete evaporation (table 2). It was shown that the film deposited on mica substrate and annealed possesses larger average area of grains than for the unannealed ones deposited on polyimide substrate. In addition, the annealing leads to decrease of the number of grains for about two times from 69 to 37 in the studied films accompanied by increase of average grain areas that are in a good agreement with [15, 16, 33].

Table 2. The average grain area <S> and areas of the grains of various sizes S1 - S4 determined from the images of surface morphology (0001) in the unannealed (sample 1) and the annealed (sample 4) p-$Bi_{0.5}Sb_{1.5}Te_3$ films.

| № in table 1 | <S>, $\mu m^2$ | S1, $\mu m^2$, % | S2 $\mu m^2$, % | S3 $\mu m^2$, % | S4 $\mu m^2$, % |
|---|---|---|---|---|---|
| 1 | 0.135 | 0.001 45 % | 0.002 - 0.008 43% | 0.01 - 0.05 10% | 2.6 - 8.7 2% |
| 4 | 1.625 | 0.002 - 0.077 72 % | 0.15 - 0.92 18 % | (9.5 - 41.25) 10 % | |

In the formed films, the thermoelectric properties depending on temperature were measured using Physical Property Measurement System (PPMS) Thermal Transport Option experimental setup.

**4. Thermal conductivity**

Total thermal conductivity of the p-$Bi_{0.5}Sb_{1.5}Te_3$ and p-$Bi_2Te_3$ films are denoted as: $\kappa = \kappa_L + \kappa_e$, where $\kappa_L$ and $\kappa_e$ are the crystal lattice and the electronic thermal conductivities, respectively. Here $\kappa_e = L(r, \eta) \sigma T$, and $L(r, \eta)$ is the Lorentz number, which was calculated using the model of the energy spectrum with isotropic scattering of charge carriers in the relaxation time approximation: $\tau = \tau_0 E^r$, the $\tau_0$ is a constant that does not depend on energy E, r is the current value of the effective scattering parameter $r_{eff}$ [34]. The effective scattering parameter $r_{eff}$ and the reduced Fermi level η were determined by the least squares method from the experimental values of the thermoelectric power coefficient α(r, η) (1) for the studied films and the degeneracy parameter $β_d$(r, η) (2), which was calculated within the framework of the many-valley model of the energy spectrum in accordance with [34, 35].

$$\alpha = \frac{k}{e}\left[\frac{(r+5/2)F_{r+3/2}(\eta)}{(r+3/2)F_{r+1/2}(\eta)} - \eta\right] \quad (1)$$

$$\beta_d(r,\eta) = \frac{(2r+3/2)^2 F_{2r+1/2}^2(\eta)}{(r+3/2)(3r+3/2)F_{r+1/2}(\eta)F_{3r+1/2}(\eta)} \quad (2)$$

The expressions α(r, η) and $β_d$(r, η) include the Fermi functions $F_{r+n}$(η), where n = 0.5, 1.5, 2.5 (figure 5):

$$F_s(\eta) = \int_0^\infty \frac{x^s}{e^{x-\eta}+1} dx \quad (3)$$

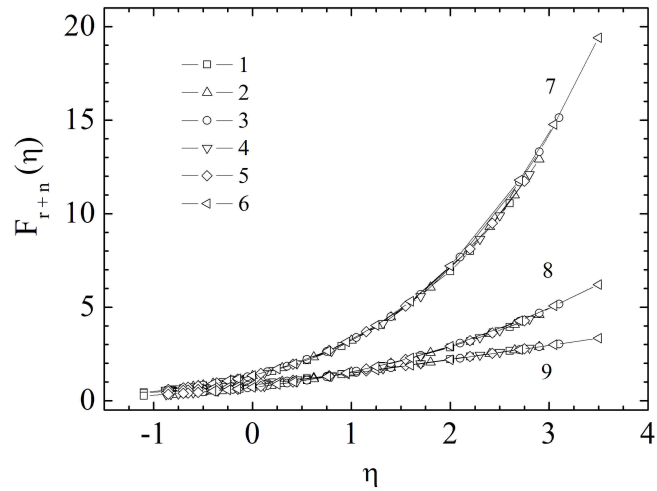

**Figure 5**. Fermi functions $F_{r+n}$(η) for the p-$Bi_{0.5}Sb_{1.5}Te_3$ (1 - 4) and p-$Bi_2Te_3$ (5, 6) films, where n: 7 - 2.5, 8 - 1.5, 9 - 0.5. Sample numbers in figure and subsequent figures correspond to Table 1.

The Lorentz number L(r, η) (4) in calculation of $κ_e$ was obtained using the data shown in figures 5 - 7:

$$L = \left(\frac{k}{e}\right)^2 \left(\frac{(r+7/2)F_{r+5/2}(\eta)}{(r+3/2)F_{r+1/2}(\eta)} - \frac{(r+5/2)^2 F_{r+3/2}^2(\eta)}{(r+3/2)^2 F_{r+1/2}^2(\eta)}\right) \quad (4)$$

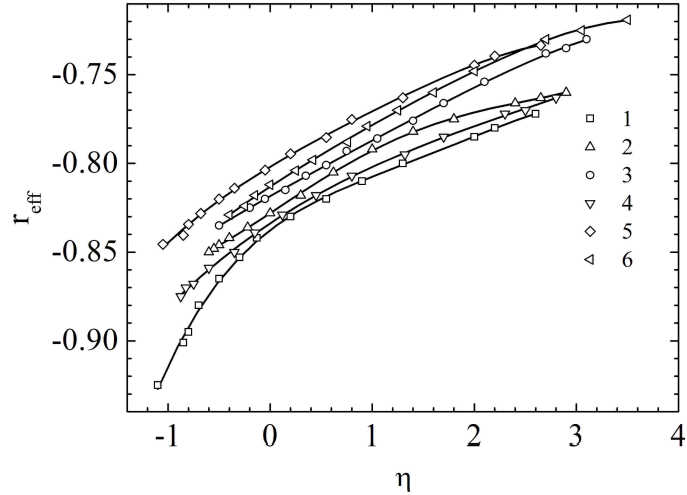

**Figure 6**. The dependences of the effective scattering parameter of the charge carriers ($r_{eff}$) on the reduced Fermi level η in the p-$Bi_{0.5}Sb_{1.5}Te_3$ (1 - 4) and p-$Bi_2Te_3$ (5, 6) films.

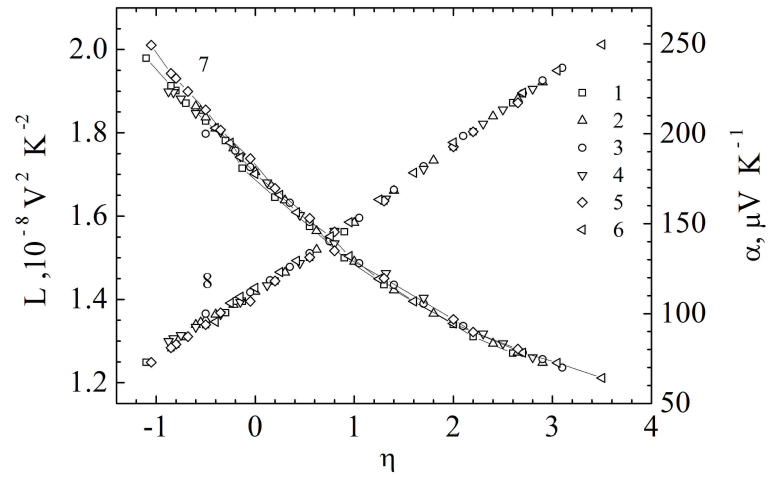

**Figure 7**. The dependences of the thermoelectric power coefficient α (points 1-6 on curve 7) and the Lorentz number L (point 1-6 on curve 8) reduced Fermi level η in the p-$Bi_{0.5}Sb_{1.5}Te_3$ (1 - 4) and p-$Bi_2Te_3$ (5, 6) films.

The dependences $r_{eff}(η)$ and α(η) (figures 6, 7) define the variation ranges of the reduced Fermi level η and the $r_{eff}$ parameter corresponding with the experimental values of the thermopower coefficient α in the temperature range of 40-300 K. The obtained $r_{eff}$ parameter (figure 6) differs from the value of r = - 0.5, specific for the acoustic scattering mechanism, but in the bulk thermoelectrics $|r_{eff}|$ is less than in the films [34]. Therefore, such increase of the $|r_{eff}|$ determines the enhancement of the relaxation time energy dependence in the films of TIs [12].

Temperature dependences of the total thermal conductivity in the films of p-$Bi_{0.5}Sb_{1.5}Te_3$ and p-$Bi_2Te_3$ were measured on samples with thickness of about 1 μm. The effects associated with the surface states of the Dirac fermions can be observed not only in ultra-thin samples owing to large quantum phase coherence length $l_\phi$ related to the inelastic electron scattering processes. The value of $l_\phi$ is usually significantly higher than the electron mean free path $l_F$, which allows to reveal the presence of the topological surface states by transport properties studies in both the nanometer thick films [4, 36 - 38] and also in the films of submicron thicknesses [12]. Obtained $r_{eff}(\eta)$ (figure 6) demonstrate an enhance of the energy dependence of the relaxation time that confirm the possibility of investigation of TSS in submicron thickness films.

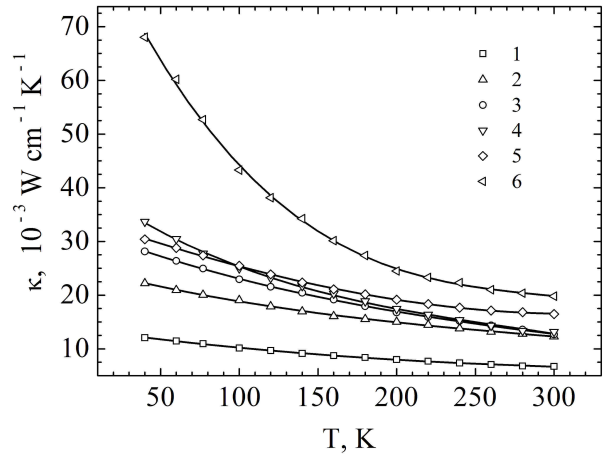

**Figure 8.** The temperature dependences of the total thermal conductivity κ in the p-$Bi_{0.5}Sb_{1.5}Te_3$ (1 - 4) and p-$Bi_2Te_3$ (5, 6) films.

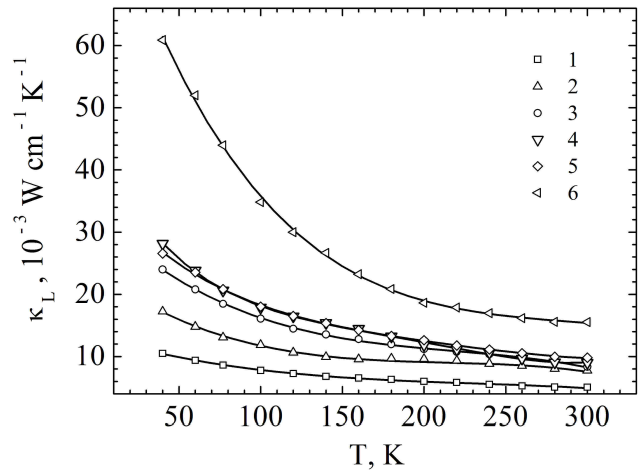

**Figure 9.** Temperature dependences of the lattice thermal conductivity $\kappa_L$ in the p-$Bi_{0.5}Sb_{1.5}Te_3$ (1 - 4) and p-$Bi_2Te_3$ (5, 6) films.

The temperature dependences of the total $\kappa$, lattice $\kappa_L$ and electron $\kappa_e$ thermal conductivities of the p-$Bi_{0.5}Sb_{1.5}Te_3$ and p-$Bi_2Te_3$ films shown in figures 8 - 10 demonstrate that these values depend on the composition, film deposition method and subsequent thermal treatment. It should be noted that the calculations of the electronic thermal conductivity $\kappa_e$ were carried out taking into account the parameter $r_{eff}(\eta)$ (figure 6). A significant decrease in $\kappa$ and $\kappa_L$ accompanied by the weakening of temperature dependences of $\kappa(T)$ and $\kappa_L(T)$ were observed in the p-$Bi_{0.5}Sb_{1.5}Te_3$ solid solution (figures 8, 9, samples 1, 2) deposited by the discrete evaporation on polyimide substrate. While, the largest decrease in $\kappa$ and $\kappa_L$ in the whole studied temperature range was obtained in the films, which were not subjected to heat treatment (figures 8, 9, sample 1).

The values of $\kappa$ and $\kappa_L$ were increased and the dependences of $\kappa(T)$ and $\kappa_L(T)$ were enhanced in the low-temperature range in the p-$Bi_{0.5}Sb_{1.5}Te_3$ films, deposited by discrete evaporation method on the mica substrates followed by annealing in the Ar atmosphere (figures 8, 9, sample 3). Further increase in $\kappa$, $\kappa_L$ and rise of $\kappa(T)$, $\kappa_L(T)$ were observed in the unannealed p-$Bi_{0.5}Sb_{1.5}Te_3$ films, formed by thermal evaporation on polyimide substrate (figures 8, 9, sample 4). In the unannealed p-$Bi_2Te_3$ films deposited on the polyimide substrate obtained by discrete evaporation, the values of $\kappa$ и $\kappa_L$ are significantly higher (figures 9, 10, sample 5) than in the p-$Bi_{0.5}Sb_{1.5}Te_3$ solid solution (figures 8, 9, sample 1). The largest values of $\kappa$ and $\kappa_L$ in the p-$Bi_2Te_3$ films were obtained by thermal evaporation (figures 8, 9, sample 6). The total $\kappa$ and lattice $\kappa_L$ thermal conductivity values (figures 8 and 9) are in good agreement with the data for the $Bi_2Te_3$ films obtained by mechanical exfoliation [33], for the films deposited onto polyimide substrates by thermal evaporation [26], and for nanocrystalline composite p-$Bi_{0.52}Sb_{1.48}Te_3$ solid solutions [16].

The considered dependences of $\kappa(T)$ and $\kappa_L(T)$ show that the optimization of technological parameters affects the intensity of the scattering of long wavelength phonons on the grain interfaces in the films and allows to significantly reduce the values of $\kappa$ and $\kappa_L$ at temperatures above the Debye temperature of $T_D = 145$ K in the unannealed p-$Bi_{0.5}Sb_{1.5}Te_3$ films, obtained by discrete evaporation on the polyimide substrate. The decrease in $\kappa$, $\kappa_L$ and the weakening of the $\kappa(T)$ and $\kappa_L(T)$ at temperatures below 100 K is determined by scattering of phonons on intrinsic antisite defects of $Bi_{Te}$ and impurity defects originated during formation of the p-$Bi_{0.5}Sb_{1.5}Te_3$ films. In addition, in this films with low values of $\kappa$, $\kappa_L$, the thermoelectric power coefficient $\alpha$ increases (figures 8, 9, samples 1, 4, table 1) due to the effect of energy filtering [9,10].

As shown in figures 1, 3, 8 and 9, a correlation was established between thermal conductivity in the films p-$Bi_{0.5}Sb_{1.5}Te_3$ obtained by discrete evaporation and the images of surface morphology (0001). In the unannealed films p-$Bi_{0.5}Sb_{1.5}Te_3$ (figures 8, 9, curves 1), the average heights $R_a$ and the root mean square deviations of the heights $R_q$ of nanofragments, the area and number of grains on the (0001) surface morphology images are increased, and the values of the thermal conductivities of $\kappa$ and $\kappa_L$ are reduced compared to the annealed films (figures 8, 9, curves 2, 4).

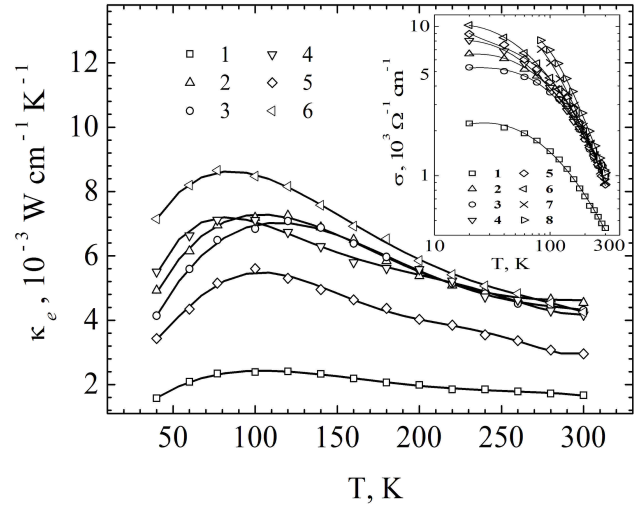

**Figure 10.** The temperature dependences of the electronic thermal conductivity $\kappa_e$ and the electrical conductivity $\sigma$ (inset) in the p-$Bi_{0.5}Sb_{1.5}Te_3$ (*1 - 4*) and p-$Bi_2Te_3$ (*5, 6*) films. The designations of the samples (1 - 6) correspond to the table 1; 7, 8 are the bulk samples of the p-$Bi_{0.5}Sb_{1.5}Te_3$ solid solution. In the films of p-$Bi_{0.5}Sb_{1.5}Te_3$ (*1 - 3*, inset) the angular coefficients of $d\ln\sigma/d\ln T = 0.1$ in the temperature interval T = (40 - 80) K and (1 - 1.2) in the interval T= (100 - 300) K.

Temperature dependences of the electronic thermal conductivity $\kappa_e$ are determined by the corresponding electrical conductivity $\sigma(T)$, (figure 10, inset), which includes the contribution of both the metallic surface and the bulk electrical conductivity of TI films [25, 39, 40]. The variation of the technological parameters of the films formation allows to reduce the bulk conductivity by optimization of the scattering of electrons on intrinsic antisite defects of $Bi_{Te}$ and impurity substitution defects of Sb → Bi in the p-$Bi_{0.5}Sb_{1.5}Te_3$ solid solution. In the p-$Bi_{0.5}Sb_{1.5}Te_3$ in low temperature range, an enhance of $\kappa_e$ with increase of the temperature from 40 to 80 K was observed (figure 10, curves 1 - 3, inset) and a significant decrease in the angular coefficients of $d\ln\sigma/d\ln T$ was obtained compared with the interval of 100-300 K. Hence the revealed feature of $\kappa_e(T)$ at low temperatures is explained by the enhancement of the energy dependence of relaxation time $\tau$ due to the rise of $|r_{eff}|$, which was used in the Lorentz number calculations.

## 5. Conclusion

The investigations of the thermal conductivity of the p-$Bi_{0.5}Sb_{1.5}Te_3$ and p-$Bi_2Te_3$ films of TIs formed by discrete and thermal evaporation techniques were carried out. It is shown that the total $\kappa$, lattice $\kappa_L$ and electronic $\kappa_e$ thermal conductivities depend not just on the composition, optimization of the film formation parameters but also on the influence of the energy dependence of the relaxation time. In addition, the $\kappa_e$ values were obtained taking into account the effective scattering parameter of $r_{eff}$ in the Lorentz number L(r, η) calculation. A significant decrease in the κ, $κ_L$ and $κ_e$ values were obtained in the unannealed films of the p-$Bi_{0.5}Sb_{1.5}Te_3$ solid solution formed by discrete evaporation on polyimide substrates.

The reduction in the crystal lattice thermal conductivity $κ_L$ in TIs films is associated with the enhancement of the energy dependence of the mean free path of phonons that leads to the intensive scattering of long-wavelength phonons on the grain interfaces at temperatures near and above $T_D$ up to room temperature, where the contribution of phonon-phonon scattering increases. In the range of low temperatures, the main reason of decrease of $κ_L$ is the scattering of short-wavelength phonons both on intrinsic point defects and on impurities. Furthermore, an additional decrease in $κ_L$ occurs due to scattering of phonons on the interfaces of the Te(1) layers of the van der Waals gap. The reduction of $κ_e$ is explained by decrease in bulk and surface electrical conductivity in the films caused by electron scattering on the intrinsic antisite and impurity point defects.

The effect of heat treatment on the relief properties of the interlayer surface (0001) studied by AFM and on the variation of the thermal conductivity $κ_L$ in the p-$Bi_{0.5}Sb_{1.5}Te_3$ films were obtained. It is shown that the total κ and the crystal lattice $κ_L$ thermal conductivities is reduced, but the number of grains on the surface (0001) and the nanofragments parameters (the average heights $R_a$ and the root mean square deviations of the heights $R_q$) are increased in the unannealed films compared with the annealed ones.

**Acknowledgements**
This study was financially supported by Russian Foundation for Basic Research Project No. 20-08-00464.

**Conflict of Interest**
The authors declare that they have no conflicts of interest.

**Data availability statement**
The data that support the findings of this study are available from the corresponding author upon reasonable request.

L N Lukyanova https://orcid.org/0000-0002-7639-9383
Yu A Boikov https://orcid.org/0000-0002-8550-2856
O A Usov https://orcid.org/0000-0001-9765-4999
V A Danilov https://orcid.org/0000-0001-7102-8551
I V Makarenko https://orcid.org/0000-0001-9661-8646